\documentclass[aps,preprint,nofootinbib,groupedaddress]{revtex4}
\usepackage{graphicx}

\newcommand{\be}{\begin{eqnarray}}
\newcommand{\ee}{\end{eqnarray}}
\newcommand{\la}{\langle}
\newcommand{\ra}{\rangle}
\newcommand{\ds}[1]{
#1{\hskip-2.0mm}/
}
\begin{document}
\preprint{ECT*-04-14}

\title{Are There Diquarks in the Nucleon?}

\author{M.~Cristoforetti$^{a\,b}$} 
\author{P.~Faccioli$^{a\,c}$}
\author{G.~Ripka$^{a\,d}$}
\author{M.~Traini$^{a\, b\, c}$}
\email{cristof@ect.it, faccioli@ect.it, ripka@cea.fr, traini@science.unitn.it}

\address{$^a$ E.C.T.$^\star$, 
286 Strada delle Tabarelle, Villazzano (Trento), I-38050 Italy}
\address{$^b$ Dipartimento di Fisica, Universit\`a degli Studi di Trento, 
Via Sommarive 15, Povo (Trento), I-38050 Italy}
\address{$^c$ Istituto Nazionale Fisica Nucleare (I.N.F.N.), 
Gruppo Collegato di Trento}
\address{$^d$ Service de Physique Theorique,
Centre d'Etudes de Saclay, F-91191, Giffur-Yvette Cedex, France.}
\date{\today}

\begin{abstract}
This work is devoted to the study of diquark correlations inside
 the nucleon.
We analyze some matrix elements
which encode information about the non-perturbative forces, 
in different color anti-triplet diquark channels.
We suggest a  lattice calculation to  check
the quark-diquark picture and clarify the role of instanton-mediated
interactions.
We study in detail the physical properties of the $0^+$ 
diquark, using the  Random Instanton Liquid Model. 
We find that instanton forces are sufficiently strong to form a diquark
bound-state, with a mass of $\sim~500$~MeV, which is compatible with earlier 
estimates. 
We also compute its electro-magnetic form factor and
find that the diquark is a broad object, with a size comparable 
with that of the proton.
\end{abstract}
\maketitle

\section{Introduction}

A complete description of the strong interaction between quarks 
is available only at asymptotically short distances, where the QCD
becomes weakly coupled.
At the perturbative level, it is well known that two quarks strongly
attract each other
if they are in a  $J^P=0^+$ and color anti-triplet state~\cite{ptdiquark1,
ptdiquark2}. 
The question whether a particularly strong attraction in this channel 
is present  also at intermediate and large distances -where the theory
is strongly coupled- has long been debated. 
From a phenomenological point of view, the notion of strongly correlated,
quasi-bound diquarks has been widely used to describe low-$q^2$ 
processes in hadrons~( for early studies see \cite{earlydiquarks1}, for
a review see~\cite{earlydiquarks2}, for further references  see also 
\cite{earlydiquarks3}). 
For example, it has been observed that the anti-triplet
scalar diquarks play an important role in non-leptonic 
weak decays of both hyperons and kaons~\cite{stech,stech2,delta12}. 

The problem of understanding  the role of diquark correlations in 
QCD has recently 
become particularly important, after several experimental groups
have reported evidence for the first truly exotic state, 
the $\Theta^+$  pentaquark~\cite{exppropentaquark}. 
The existence of such a state was 
originally predicted by Diakonov, Petrov and Polyakov
using a soliton model~\cite{pentaquark} and several experiments are 
being performed to confirm 
this important discovery. There are also on-going theoretical inverstigations
based on lattice QCD simulations~(see for example~\cite{saza} and
references therein).

The claims of the observation of the pentaquark have triggered a huge
theoretical activity, aiming to identify possible dynamical mechanisms
which could explain the existence and the 
extremely narrow width of this resonance.
Jaffe and Wilczek~\cite{jw} have suggested that
several important properties of the hadronic spectrum, including the 
 existence of a pentaquark anti-decuplet, 
could be explained by assuming 
that the non-perturbative quark-quark interaction is particularly 
attractive in the  anti-triplet $0^+$ channel, leading 
to strongly correlated scalar $u-d$~diquarks, inside hadrons.

The implications of the diquark hypothesis on the 
structure and the decay of the pentaquark has been analyzed in a number of 
recent works~(for an incomplete list see \cite{incomplete}).
At the same time, the Jaffe-Wilczek analysis 
has opened a related discussion concerning the possible
non-perturbative mechanisms in QCD, which can lead to a strong 
attraction in the scalar anti-triplet channel. 
Shuryak and Zahed have performed an analysis of the hadronic spectrum
assuming that diquark correlations are 
induced by instantons~\cite{diquarkSHU}. In addition to diquarks,  
Kochelev, Lee and Vento have also suggested the existence of instanton-induced 
tri-quarks~\cite{diquarkVENT}.
It should be mentioned all models in which the quarks 
in the pentaquark are strongly correlated contrast with the 
original mean-field picture proposed by Diakonov, Petrov and Polyakov,
in which  two-body correlations are neglected~\cite{pentaquark}.

Even though the experimental evidence for the $\Theta^+$ 
pentaquark is still a matter of debate, understanding if strong
correlations in the $0^+$ anti-triplet diquark channel are present in the
vacuum and in hadrons is an important open problem in QCD.
At present, Lattice Gauge Theory  represents 
the only available  framework to 
perform ab-initio non-perturbative calculations in QCD.
It is therefore natural to attempt to use it
to address this fundamental question.
In this paper, we take a first step in this direction. 
We identify and study a set of lattice-calculable matrix-elements, which 
are very sensitive to the quark-quark  interaction
in different diquark channels.
In order to show that these matrix elements can be used to gain insight
on diquark correlations inside hadrons, we compute them
using  three different phenomenological models: a naive 
$SU(6)$ Non-Relativistic Quark Model, in which dynamical diquark 
correlations are absent, 
the Random Instanton Liquid Model, where scalar 
diquarks are strongly favored by the 't~Hooft interaction, 
and a Chiral Soliton Model, where the 't~Hooft interaction is
treated at the mean-field level.
We show that these  models lead to very different predictions. 
We also  identify a lattice 
calculation which can check if the strong attraction in the scalar
anti-triplet diquark channel can be mediated by instantons, 
as in the model of Shuryak and Zahed. 

In the second part of this work, we study the physical properties of
the diquarks induced by instanton-mediated forces. 
We provide unambiguous evidence that the 't~Hooft interaction
leads to a $0^+$ bound-state of about 500~MeV. This result
agrees with the early estimate by Shuryak and collaborators~\cite{baryonRILM}. 
We also compute the electro-magnetic form factor of such a bound-state,
finding that its size is comparable with that of the proton. This result
suggests that diquarks cannot at all be modeled as point-like objects.

The paper is organized as follows. In section \ref{densities}, we 
define the relevant matrix elements and explain why these quantities
 can be used to study diquark correlations, in different diquark channels. 
The matrix elements are computed in different models
in section~\ref{models} and the results are analyzed in 
section~\ref{discussion}. The diquark mass and size are computed in the Random
Instanton Liquid Model,  
in section \ref{sizemass}. All the results and the conclusions 
of this work are summarized in section~\ref{conclusions}.

\section{Diquark densities in the proton}
\label{densities}

Much of the theoretical studies of the internal structure of hadrons
focus on observables related to  one-body local operators.
These matrix elements allow to access information on the hadron
internal wave-function, but are 
only indirectly
sensitive to the Dirac and flavor
structure of the non-perturbative quark-quark interaction.
For example, models of the proton
with drastically different diquark content typically
give comparable charge radii.

Much more direct insight on quark-quark correlations inside
hadrons 
can be learned by focusing on {\it two-body} local density operators, which 
simultaneously probe the position and the discrete quantum numbers of two 
quarks. 
In this view, we consider a set of
four-field operators (which we shall refer to as {\it diquark densities}), 
in which two quarks with quantum
numbers of a scalar, vector, axial-vector and pseudo-scalar color anti-triplet diquark are destroyed and re-emitted, in the same point.

We work with Euclidean 
space-time\footnote{In our notation $x=({\bf r},x_4)$.}, and we define:
\be
\label{D}
D_\Gamma({\bf r})&=& 
F_\Gamma^{a\,\dagger}({\bf  r},0) F^a_{\Gamma}
({\bf r},0),\\
F^a_{\Gamma}(x)&=& 
\epsilon^{a b c} u^b(x)\,C\,\Gamma\,d^c(x),
\label{F}
\ee
where $C=i\,\gamma_2\,\gamma_4$ is the charge-conjugation operator, 
$\Gamma\in\{ 1,\gamma_5, \gamma_\mu,\gamma_\mu~\gamma_5, \sigma_{\mu,\nu}\}$ 
and $a,b,c$ are  color indexes
The operator $F^a_\Gamma(x)$ absorbs two quarks in a point with quantum 
numbers of a color anti-triplet diquark.
For example, $F^a_{\gamma_3}(x)$ annihilates an axial-vector diquark, 
while $F^a_{\gamma_5}(x)$ 
absorbs scalar diquarks. Hence, the matrix element:
\be
\label{rho}
\rho_\Gamma({\bf r})=\langle P| D_\Gamma ({\bf r}) 
|P \rangle
\ee
measures the probability amplitude to find at time $t=0$ two quarks 
at the point ${\bf r }$ in the proton, in a anti-triplet color 
state and with quantum numbers specified by $\Gamma$.

The choice of  $\Gamma$  in (\ref{F}) determines the transformation property
of the $F_\Gamma^a(x)$ operator under exchanges of the flavor indexes.
Acting on the vacuum, the operator 
$F_{\Gamma}^{a\,\dagger }\left({\bf r}, 0\right) $
creates at time $t=0$ the diquark state: 
\begin{equation}
F_{\Gamma}^{a\,\dagger }
\left( {\bf r},0 \right) \left| 0\right\rangle =\varepsilon^{abc}\left( \Gamma ^{\dagger }C\right) _{s\,t}
\left| {\bf r}\, s\,d\,b; {\bf r}\,t\,u\,c\right\rangle,
\label{fga}
\end{equation}
where $s,t$ are spinor indexes, $a,b,c$ are color indexes, $u, d$ are quark
flavors.
This state is symmetric under the exchange of the coordinates ${\bf r}$ and
antisymmetric under the exchange of color.
Under the exchange of the Dirac indexes, the state is either symmetric or
anti-symmetric, depending on $\Gamma $, as shown in table~\ref{tablesymm}: 
\begin{table}
\begin{tabular}{|l|l|l|l|l|l|}
\hline
$\Gamma $ & space & Dirac & flavor & color & total \\ \hline
$1$ & S & A & A $\left( T=0\right) $ & A & A \\ \hline
$\gamma _{5}$ & S & A & A $\left( T=0\right) $ & A & A \\ \hline
$\gamma ^{4}$ & S & S & $S\left( T=1,M_{T}=0\right) $ & A & A \\ \hline
$\gamma ^{3}$ & S & S & S $\left( T=1,M_{T}=0\right) $ & A & A \\ \hline
$\gamma ^{4}\gamma ^{5}$ & S & A & A $\left( T=0\right) $ & A & A \\ \hline
\end{tabular}
\caption{Symmetry properties of the operators $F^a_\Gamma(x)$, under
exchange of space, color, flavor and Dirac indexes.}
\label{tablesymm}
\end{table}
for $\Gamma =\gamma _{5}$, the matrix $%
\Gamma ^{\dagger }C$ is antisymmetric and the diquark state is thus
antisymmetric under the exchange of Dirac indexes. The state is symmetric
under the exchange of space indexes and antisymmetric under the exchange of
color indexes. It is by construction antisymmetric with respect to the
exchange of all indexes (space, Dirac, flavor and color) so that it is
necessarily antisymmetric with respect to flavor indexes, meaning that it is
a flavor $T=0$ state. However, when $\Gamma =\gamma_{4}$ (~a symmetric
matrix~) the diquark state is symmetric under the exchange of Dirac indexes.
The state is thus necessarily symmetric with respect to flavor indexes,
meaning that it is a flavor $T=1$ $M_{T}=0$ state.

The comparison of the different matrix elements (\ref{rho}) can 
provide useful insight into the internal dynamics of the hadron, in particular
about the strength and the spin- and flavor- structure of 
the quark-quark interaction.
In general, we expect that the stronger are the two-body correlations inside
the hadron,
the larger are the discrepancies between a mean-field description and the 
exact calculation.
More specifically, if the quark-quark interaction is 
particularly attractive in the scalar anti-triplet diquark channel 
$-$~as it is assumed in the Jaffe-Wilczek picture~$-$
then we expect that the scalar diquark density $\rho_{\gamma_5}({\bf r})$ 
should be enhanced  with respect to a model in which the interaction is not
particularly attractive, in this channel. In fact, 
the $u$ and $d$ 
quarks would have a larger probability ``to be found'' in the same point and
be destroyed by the local operator $F_{\gamma_5}^a(x)$.

The diquark densities can also be used to study relativistic effects
in the hadron. The $F^a_1(x)$ and  
$F^a_{\gamma_4}(x)$ operators mix upper and
lower spinor components of the quark fields and vanish in 
the non-relativistic limit.
Hence the densities
$\rho_1({\bf r})$ and 
$\rho_{\gamma_4}( {\bf r})$ are sensitive only to the relativistic
components of the wave-function.
 On the contrary the operators $F^a_{\gamma_3}(x)$,
$F^a_{\gamma_5\gamma_4}(x)$, and $F^a_{\gamma_5}(x)$ remain finite 
in the non-relativistic limit.

\section{Diquark densities in different models}
\label{models}

In this section, we compute the 
predictions for the diquark densities defined in the previous section, 
using the Chiral Soliton Model, the $SU(6)$ Non-Relativistic Quark Model, 
and the Random Instanton Liquid Model. 
The Non-Relativistic Quark Model is used to mimic a scenario in 
which the quarks are very weakly {\it dynamically} correlated and do not 
form bound diquarks. In such a model, 
all spin-flavor correlations are due to the $SU(6)$ symmetry properties of
the wave-function only.
The Random Instanton Liquid Model calculation represents the opposite
scenario, in which the non-perturbative
interaction is spin-flavor dependent and is particularly attractive
in the $0^+$ anti-triplet diquark channel.
The mean-field Chiral Soliton Model can be used to mimic a 
scenario in which the 
same interaction binds the nucleon, but 
leads to negligible two-body correlations.

The reader who is not interested in the technical 
details of these calculations may skip the remaining part of this section and 
consider directly the discussion of the results presented
in section~\ref{discussion}.

\subsection{Chiral Soliton Model}

The Chiral Soliton Model~(for a detailed discussion
we refer the reader to e.g. \cite{chiralsoliton, Ripka}) 
can be used to account for the 't~Hooft interaction in the nucleon
to all orders, keeping only the leading terms in the $1/N_c$ expansion.
Hence, in the large $N_c$ limit, the Random Instanton Liquid Model and 
Chiral Soliton Model should give similar results.

As all mean-field approaches, the Chiral Soliton Model is most 
efficient for calculations  of matrix elements of one-body operators.
Indeed, the Random Instanton Liquid Model and Chiral Soliton 
Model give similar predictions 
for the proton form factors~\cite{solitonFF,myformfactors}.
On the other hand, {\it if} the instanton-induced two-body correlations are  
strong for $N_c=3$, then we expect that the Random Instanton Liquid Model and 
the Chiral Soliton Model should give significantly 
different results for matrix elements of two-body 
operators\footnote{For example, 
the Chiral Soliton Model may not be very accurate in predicting the
amplitudes for weak decays of hadrons, which are driven by a four-field
effective Hamiltonian.}, such as the diquark 
densities (\ref{rho}).

For $N_f=2$, the Chiral Soliton Model is 
derived by bosonization of the four-field 
't~Hooft contact interaction, introducing an auxiliary chiral field. 
The chirally-invariant Hamiltonian governing the dynamics of the 
quarks field $\psi$
and  the chiral fields $\sigma$ and $\pi_a$ ($a=1,2,3$) is:
\be
\label{HCSM}
{\mathcal H} 
= \int d^3{\bf r}~\psi^\dagger \left[{\bf \alpha} \cdot {\bf p}+ g 
\beta (\sigma + i\gamma_5 {\bf \pi}\cdot {\bf \tau})\right]\psi
\nonumber\\
+\int d^3{\bf r} 
\left[\frac{1}{2}(\nabla \sigma)^2+ \frac{1}{2}(\nabla {\bf \pi})^2+
\lambda (\sigma^2+{\bf \pi}^2 -\sigma_0^2)^2\right],
\ee
where $\tau_a$ are isospin Pauli matrices, $g$ and $\lambda$ are coupling
constants and $\sigma_0$ is the vacuum expectation value
of the scalar field $\sigma$.
In the large $N_c$ limit, one can apply
 the zero-loop approximation and
treat the ($\sigma, {\bf \pi}$) fields as classical.

The Hamiltonian (\ref{HCSM}) is quadratic in the quark fields
and gives rise to the
single-particle Dirac equation
\be
\label{csmqua}
\left[ -i\,{\bf \alpha \cdot  \nabla} + 
\beta (\phi +i \gamma_5 {\bf \chi}\cdot{\bf \tau})\right]
|\lambda\rangle= \epsilon_\lambda |\lambda\rangle,
\ee
where, for convenience, we have introduced dimensionless distances and fields
defined as:
\be 
{\bf x} = g\,\sigma_0 {\bf r},
\qquad
\phi = \frac{\sigma}{\sigma_0}, 
\qquad
\chi_a = \frac{\pi_a}{\sigma_0}.
\label{scaling}
\ee
The total energy, when  $N_c$ quarks are accommodated in the 
${\bf J} + {\bf I}= {\bf 0}$ valence 
orbital of energy $\epsilon_\lambda$, is 
\be
E = N_c\,g\,\sigma_0~\epsilon_\lambda +\frac{\sigma_0}{2 g}
~\int d^3~{\bf x} \left[ (\nabla \phi)^2+(\nabla \chi)^2 + \frac{\lambda}{g^4}
(\phi^2+{\bf \chi}^2-1)^2\right].
\ee
By requiring that the total energy must be stationary with respect
to infinitesimal variations of the chiral fields one obtains the equations:
\be
\label{csmphi}
\frac{-1}{g^2}\,\nabla^2\phi+ 4\frac{\lambda}{g^4}\phi~(\phi^2+{\bf \chi}^2-1)
+ N_c \langle\lambda| {\bf x} \rangle~\beta~\langle {\bf x} | 
\lambda \rangle=0,\\
\label{csmchi}
\frac{-1}{g^2}\,\nabla^2\chi_a+ 4\frac{\lambda}{g^4}\chi_a~
(\phi^2+{\bf \chi}^2-1) + N_c \langle\lambda| {\bf x} \rangle~i\beta\,\gamma_5
~\langle {\bf x} | \lambda \rangle=0,
\ee
Eq.s (\ref{csmqua}), (\ref{csmphi}) and (\ref{csmchi}) represent a 
self-consistent set of equations for the quark orbitals $|\lambda\rangle$
and the classical chiral fields.  
The chiral soliton state is a solution of such equations, 
in which the
chiral fields are assumed to be 
time-independent and have a hedgehog shape:
\be
\label{solstate}
\phi(x)=\phi({\bf x}),\qquad
\chi_a(x)=\hat{{\bf x}}_a \chi({\bf x}).
\ee
In the chiral soliton model of Diakonov  and Petrov~\cite{chiralsoliton},
the  fields $\sigma$ and $\pi_a$ are restricted to the chiral circle, 
$\sigma^2+ {\bf \pi}^2 = \sigma_0^2$.
With the  ansatz (\ref{solstate}), the Dirac equation (\ref{csmqua}) 
generates a bound-state composed of valence and Dirac-sea quarks:
\be
|\Phi\rangle=\prod_{\mu\in valence}~a^\dagger_\mu~\prod_{\lambda\in sea} 
a^\dagger_\lambda~ |0\rangle.
\ee
The first product of creation operators excites three valence 
quarks of different
colors in a $s$-state wave-function in the form:
\be
\label{val}
\langle {\bf x}~u| \lambda \rangle =
\frac{1}{\sqrt{2}}\left(
\begin{array}{c}
i  \frac{F({\bf x)}}{|{\bf x}|}~|\downarrow~\rangle \\
-\frac{G({\bf x})}{|{\bf x}|}{\bf \sigma \cdot \hat{x}}~|\downarrow~\rangle
\end{array}
\right),\qquad
\langle {\bf x}~d| \lambda \rangle =
\frac{1}{\sqrt{2}}\left(
\begin{array}{c}
-i  \frac{F({\bf x)}}{|{\bf x}|}~|\uparrow~\rangle \\
\frac{G({\bf x})}{|{\bf x}|}{\bf \sigma \cdot \hat{x}}~|\uparrow~\rangle
\end{array}
\right)
\ee
The radial functions $F({\bf x})$ and $G({\bf x})$ 
are normalized according to 
\be
\int_0^\infty d^3 {\bf x}~[~F^2({\bf x})+G^2({\bf x})~]=1,
\ee
and can  be computed numerically.

The second product of creation operators in (\ref{solstate}) generates
the contribution of quarks in the Dirac sea.
In this work, we shall neglect such a contribution and retain only the
valence part of the soliton wave-function.

After having solved for the valence component of the  
soliton wave-function $|\Phi\rangle$, 
it is straightforward to compute the different diquark densities:
\begin{equation}
\rho _{\Gamma}\left( {\bf r}\right) =\left\langle \Phi \left| 
\,F_{\Gamma}^{a\,\dagger}\left( {\bf r},0\right) F_\Gamma^{a}\left( {\bf r},0\right) \right| 
\Phi \right\rangle,
\qquad
\Gamma= 1, i\gamma_5, \gamma_4, \gamma_3, \gamma_4\gamma_5.
\end{equation}
We begin by contracting the color index $a$.
The diquark density becomes: 
\be
\rho_\Gamma\left( {\bf r}\right) =\left( \Gamma C\right) _{s_{1}s_{2}}\left(
C\Gamma\right) _{s_{3}s_{4}}\left\langle \Phi \left|
d_{s_{1}}^{c_{3}\,\dagger }u_{s_{2}}^{c_{2}\,\dagger
}u_{s_{3}}^{c_{2}}d_{s_{4}}^{c_{3}}-d_{s_{1}}^{c_{3}\,\dagger
}u_{s_{2}}^{c_{2}\,\dagger }u_{s_{3}}^{c_{3}}d_{s_{4}}^{c_{2}}\right| \Phi
\right\rangle,
\ee
where $s_1,s_2,s_4,s_4=1,..,4$ are Dirac indexes and $c_2,c_3=1,...,3$ 
are color indexes.
By applying Wick's theorem we get:
\be
&\mbox{}&
\left\langle \Phi \left| d_{s_{1}}^{c_{3}\,\dagger }u_{s_{2}}^{c_{2}\,\dagger}u_{s_{3}}^{c_{2}}d_{s_{4}}^{c_{3}}-d_{s_{1}}^{c_{3}\,\dagger
}u_{s_{2}}^{c_{2}\,\dagger }u_{s_{3}}^{c_{3}}d_{s_{4}}^{c_{2}}\right| \Phi
\right\rangle
=\nonumber\\ &=&
N_{c}^{2}\sum_{\lambda \mu \in D}\left\langle \lambda \left| {\bf x}%
s_{1}d\right. \right\rangle \left\langle {\bf x}s_{4}d\left| \lambda \right.
\right\rangle \left\langle \mu \left| {\bf x}s_{2}u\right. \right\rangle
\left\langle {\bf x}s_{3}u\left| \mu \right. \right\rangle
\nonumber\\ &+&
N_{c}^{2}\sum_{\lambda \mu \in D}\left\langle \lambda \left| {\bf x}%
s_{1}d\right. \right\rangle \left\langle {\bf x}s_{3}u\left| \lambda \right.
\right\rangle \left\langle \mu \left| {\bf x}s_{2}u\right. \right\rangle
\left\langle {\bf x}s_{4}d\left| \mu \right. \right\rangle 
\nonumber\\ &-&
N_{c}\sum_{\lambda \mu \in D}\left\langle \lambda \left| {\bf x}%
s_{1}d\right. \right\rangle \left\langle {\bf x}s_{3}u\left| \lambda \right.
\right\rangle \left\langle \mu \left| {\bf x}s_{2}u\right. \right\rangle
\left\langle {\bf x}s_{4}d\left| \mu \right. \right\rangle
\nonumber\\ &-&
N_{c}\sum_{\lambda \mu \in D}\left\langle \lambda \left| {\bf x}%
s_{1}d\right. \right\rangle \left\langle {\bf x}s_{4}d\left| \lambda \right.
\right\rangle \left\langle \mu \left| {\bf x}s_{2}u\right. \right\rangle
\left\langle {\bf x}s_{3}u\left| \mu \right. \right\rangle.
\ee
Substituting the explicit expression for the valence orbits~(\ref{val}), 
and performing the summation over the spinor indexes, it is straightforward
to obtain the result (choosing ${\bf x}$ along the $\hat{z}$ direction):
\be
\label{resSOLi}
\rho_1({\bf r}) &=&0,\\
\rho_{\gamma_5}({\bf r})&=& N_c (N_c-1)\left(\frac{F^2(g\,\sigma_0\,{\bf r}) +
G^2(g\,\sigma_0\,{\bf r})}{\sqrt{2}\,(g\,\sigma_0\,{\bf r})^2}\right)^2,\\
\rho_{\gamma_4}({\bf r})&=& 2 N_c 
(N_c-1)\left(\frac{F(g\,\sigma_0\,{\bf r})~G(g\,\sigma_0\,{\bf r})}
{(g\,\sigma_0\,{\bf r})^2}\right)^2,\label{resSOL4}\\
\rho_{\gamma_3}({\bf r})&=& N_c (N_c-1)\left(\frac{F^2(g\,\sigma_0\,{\bf r})-
G^2(g\,\sigma_0\,{\bf r})}{\sqrt{2}\,(g\,\sigma_0\,{\bf r})^2}\right)^2,\\
\label{resSOLf}
\rho_{\gamma_4\gamma_5}({\bf r}) &=& N_c (N_c-1)
\left(\frac{F^2(g\,\sigma_0\,{\bf r})-
G^2(g\,\sigma_0\,{\bf r})}{\sqrt{2}\,(g\,\sigma_0\,{\bf r})^2}\right)^2, 
\ee
where we have used 
the relationship between ${\bf x}$ and ${\bf r}$  given by~(\ref{scaling}).

These results correspond to the different diquark densities in the soliton
state, which is neither an eigenstate of the angular momentum nor an eigenstate
of the isospin operator. Instead, it is characterized by its (vanishing) 
eigenvalue of the grand-spin ${\bf G}={\bf J} + {\bf I}$ operator.
In order to make contact with the proton, one has to perform a projection onto
a state with $(T=1/2, M_T=1/2)$.
The operators $F_{\Gamma}^{a\,\dagger }\left( x\right) $ are all color zero
and flavor zero operators, except for the operators $F_{\gamma _{4}}^{a\,\dagger}\left( x\right) $ and $F_{\gamma_{3}}^{a\,\dagger }\left( x\right) $.\ Thus
the operators $D_{\Gamma }\left( {\bf r},0 \right) =
\,F_{\Gamma}^{a\,\dagger}
\left( {\bf r},0\right) F_{\Gamma}^a\left( {\bf r},0\right) $ with $\Gamma
=\left( 1,\gamma _{5},\gamma_4\gamma _{5}\right) $ are flavor zero
operators, so that their expectation value in the nucleon state is equal to
their expectation value in the soliton state.

This, however, is not true for the operators $F_{\gamma_4}^{a\,\dagger}\left( x\right) $ and $F_{\gamma _{3}}^{a\,\dagger}\left( x\right) $ which
are flavor $T=1,\;M_{T}=0$ operators.\ Indeed, acting at time
$t=0$ on a flavor zero state 
$\left| 0\right\rangle $ (which is \emph{not} the soliton state), the
operator $F_{\gamma_4}^{a\,\dagger}$ produces the state 
\begin{equation}
F_{\gamma_4}^{a\,\dagger }\left( {\bf r}, 0\right) \left| 0\right\rangle
=\varepsilon^{abc}\left( \gamma_4\,C\right) _{st}\left|
{\bf r} \,s\,d\,b; {\bf r}\,t\,u\,c\right\rangle,
\end{equation}
which is symmetric under the exchange of the flavor indexes. Similarly, the
hermitian conjugate operator $F_{\gamma_4}^a$ produces the
state: 
\begin{equation}
F_{\gamma_4}^a\left({\bf r}, 0\right) \left| 0\right\rangle 
=\varepsilon^{abc}\left( C\gamma_4\right)
_{ts}\left|{\bf r}\,t\,d\,b; {\bf r}\,s\,u\,c\right\rangle,
\end{equation}
which is a flavor $T=1,M_{T}=0$ state. Therefore the operator 
$D_{\gamma_4}({\bf r})~=~F_{\gamma_4}^{a\,\dagger}({\bf r}, 0)~
F_{\gamma_{4}}^a({\bf r}, 0) $ behaves,
under flavor rotations as the following mixture of $T=0$ and $T=2$
operators: 
\begin{equation}
D_{\gamma_4}({\bf r})=%
\frac{1}{\sqrt{3}}~D_{\gamma_4}^{T=0,M_{T}=0}
\left( {\bf r}\right) -\sqrt{\frac{2}{3}}
D_{\gamma_4}^{T=2,M_{T}=0}\left( {\bf r}\right),  \label{ff02}
\end{equation}
where $\frac{1}{\sqrt{3}}$ and $-\sqrt{\frac{2}{3}}$ are Clebsch-Gordon
coefficients.\ The second term, which is a flavor $T=2$ operator, has a
vanishing expectation value in the nucleon state 
which has flavor $T=\frac{1}{2}$.\ Therefore only the first term of (\ref
{ff02}) contributes to the diquark density in the proton: 
\be
\rho_{\gamma_4}({\bf r})=
\langle P | D_{\gamma_4}({\bf r}) | P \rangle 
=\frac{1}{\sqrt{3}}\langle P | 
D_{\gamma_4}^{T=0,M_{T}=0}({\bf r})| P \rangle.
\ee
We need to determine the operator $D^{\gamma_4}_{T=0,M_{T}=0}\left(
{\bf r}\right) $. 
To do this we use a simplified and more transparent notation: 
\begin{equation}
F_{\gamma_4}^{a\,\dagger }\left( x \right) =F_{du}^{a\,\dagger
}(x)=F_{ud}^{a\,\dagger }(x)\;\;\;\;F^a_{\gamma_4}\left( x \right) 
=F_{du}^a(x)=F_{ud}^a(x).
\end{equation}
Then the inverse Clebsch expansion reads: 
\be
D_{\gamma_4}^{T=0,M_{T}=0}\left( {\bf r}\right) 
=-\frac{1}{\sqrt{3}}F_{uu}^{a\,\dagger}({\bf r},0)\,
F_{uu}^a({\bf r},0)-
\frac{1}{\sqrt{3}}F_{dd}^{a\,\dagger }({\bf r},0)
F_{dd}^a({\bf r},0)+
\frac{1}{\sqrt{3}}F_{du}^{a\,\dagger }({\bf r},0)\,F^a_{du}({\bf r},0).\nonumber\\
\ee
On the right hand side, the operators $F_{uu}^{a\,\dagger }$ and $%
F_{dd}^{a\,\dagger }$ are the $T=1,M_{T}=\pm 1$ operators: 
\begin{equation}
F_{uu}^{a\,\dagger }(x)=\varepsilon^{abc}u_{s}^{c\,\dagger }\left( x\right) \left(
\gamma_4\,C\right) _{st}u_{t}^{b\,\dagger }\left( x\right)
\;\;\;\;\;F_{dd}^{a\,\dagger }(x)=\varepsilon^{abc}d_{s}^{c\,\dagger }\left(
x\right) \left( \gamma_4\,C\right) _{st}d_{t}^{b\,\dagger }\left( x\right).
\end{equation}
Since the
$D_{\gamma_4}^{T=0,M_{T}=0}\left( {\bf r} \right) $ is a flavor zero
operator, its expectation value in the proton state $\left| P
\right\rangle $ is equal to its expectation value in the soliton state $\left|
\Phi \right\rangle $.\ Therefore, the diquark density 
$\rho _{\gamma_4}\left( {\bf r}\right) $ of the proton is: 
\be
\rho _{\gamma_4}\left( {\bf r}\right)& =&
\frac{1}{\sqrt{3}}\langle
\Phi| -\frac{1}{\sqrt{3}}F_{uu}^{a\,\dagger }({\bf r},0)\,
F_{uu}^a({\bf r},0)-\frac{1}{\sqrt{3}}F_{dd}^{a\,\dagger }({\bf r},0)\,
F^a_{dd}({\bf r},0)+\frac{1}{\sqrt{3}}\,F_{du}^{a\,\dagger }({\bf r},0)\,F^a_{du}({\bf r},0)
|\Phi \rangle\nonumber\\
&=&-\frac{1}{3}~\rho_{\gamma_4}^{uu}({\bf r})-
\frac{1}{3}\rho_{\gamma_4}^{dd}({\bf r}) + 
\frac{1}{3}~\rho_{\gamma_4}^{ud}({\bf r}).
\label{rhgam0}
\ee
The densities $\rho_{\gamma_4}^{uu}({\bf r})$ and 
$\rho_{\gamma_4}^{dd}({\bf r})$ can be 
computed in complete
analogy with the $\rho^{\gamma_4}_{ud}({\bf r})$  (which leads to
Eq. (\ref{resSOL4})).
For a generic direction $\hat{{\bf r}}$ the result is
\begin{equation}
\rho_{\gamma_4}^{uu}\left( {\bf r}\right) =\rho_{\gamma_4}^{dd}
\left({\bf r}\right)
=-2N_{c}\left( N_{c}-1\right) 
\frac{F(g\,\sigma_0\,{\bf r})\,G(g\,\sigma_0\,{\bf r})}
{2(g\,\sigma_0\,{\bf r})^{2}}\left( \widehat{r}^{2}_x+%
\widehat{r}^{2}_y\right).  \label{rhouu}
\end{equation}
Since we have chosen  ${\bf r}$ along the $\hat{z}$ direction, 
we get a vanishing contribution.
So, in the Chiral Soliton Model,  the density $\rho_{\gamma_4}({\bf r})$ 
in the proton state (as opposed to the density in the soliton state) reads:
\be
\rho_{\gamma_4}({\bf r})=\frac{2}{3}\,N_c(N_c-1)
\left(\frac{G(g\,\sigma_0\,{\bf r})\,F(g\,\sigma_0\,{\bf r})}
{(g\,\sigma_0\,{\bf r})^2}\right)^2.
\label{rho4P}
\ee
Notice that  projecting onto the proton
leads to a reduction of this density by a factor 3.

We can repeat the same calculation to obtain the Chiral Soliton Model
prediction for the $\rho_{\gamma_3}({ \bf r})$ density
in the proton.
Also in this case, the contribution from the $\rho_{\gamma_3}^{uu}$ 
and $\rho_{\gamma_3}^{dd}$
densities vanish, if ${\bf r}$ is chosen along the $\hat{z}$ 
direction. The final result is 
\be
\rho_{\gamma_3}({\bf r})&=& \frac{1}{3}\,N_c (N_c-1)\left
(\frac{F^2(g\,\sigma_0\,{\bf r})-
G^2(g\,\sigma_0\,{\bf r})}{\sqrt{2}\,(g\,\sigma_0\,{\bf r})^2}\right)^2.
\label{rho3P}
\ee
Also in this channel, the projection onto the proton has generated an
extra $1/3$ factor.

In conclusion,  the Chiral Soliton Model predictions for the diquark densities
in the proton are:
\be
\label{resSOLpi}
\rho_1({\bf r}) &=&0,\\
\rho_{\gamma_5}({\bf r})&=& N_c (N_c-1)~\left(\frac{F^2(g\,\sigma_0\,{\bf r})+
G^2(g\,\sigma_0\,{\bf r})}{\sqrt{2}\,(g\,\sigma_0\,{\bf r})^2}\right)^2,\\
\rho_{\gamma_4}({\bf r})&=& \frac{2 N_c (N_c-1)}{3}~
\left(\frac{F(g\,\sigma_0\,{\bf r})~G(g\,\sigma_0\,{\bf r})}
{(g\,\sigma_0\,{\bf r})^2}\right)^2,\\
\rho_{\gamma_3}({\bf r})&=& \frac{N_c (N_c-1)}{3}~
\left(\frac{F^2(g\,\sigma_0\,{\bf r})-
G^2(g\,\sigma_0\,{\bf r})}{\sqrt{2}\,(g\,\sigma_0\,{\bf r})^2}\right)^2,\\
\label{resSOLpf}
\rho_{\gamma_4\gamma_5}({\bf r}) &=& N_c (N_c-1)
\left(\frac{F^2(g\,\sigma_0\,{\bf r})-
G^2(g\,\sigma_0\,{\bf r})}{\sqrt{2}\,(g\,\sigma_0\,{\bf r})^2}\right)^2.
\ee

\subsection{Non-Relativistic SU(6) Quark Model}

In the conventional $SU(6)$ Non-Relativistic Quark Model, the proton 
wave-function is defined as: 
\begin{eqnarray}
\label{proton}
|P_{\uparrow}\ra &=& 
\frac{1}{\sqrt{2}}~\left(\chi_{MS}\,\phi_{MS} + \chi_{MA}\,\phi_{MA}\right)\, 
\otimes |\textrm{color}\rangle ~\otimes~|\textrm{spatial}\rangle,
\end{eqnarray}
where 
\be
\chi_{MA}=\frac{1}{\sqrt{2}} 
|(\uparrow\,\downarrow -\downarrow\, \uparrow)
\uparrow \rangle,
\qquad \phi_{MA}=\frac{1}{\sqrt{2}}~|(u\,d -d\,u) u \rangle,
\ee
\be
\chi_{MS}=\frac{1}{\sqrt{6}}~
|(\uparrow\,\downarrow +\downarrow\, \uparrow)\,\uparrow 
-2 \uparrow\,\uparrow\, \downarrow~\rangle,
\qquad
\phi_{MS}=\frac{1}{\sqrt{6}}
|(u\,d + d\,u)u - 2 u\,u\,d \rangle.
\ee
The spacial (color) wave-function is totally symmetric (anti-symmetric).

The diquark density operators are constructed by expanding the field
operators in $F_\Gamma ^a(x)$ on a basis of the constituent quark
creation/annihilation operators and taking the non-relativistic limit for the
spinors. 
\begin{eqnarray}
F_\Gamma ^a(\mathbf{r},0) &=& 
\sum_{s^{\prime },s=\uparrow ,\downarrow }\widehat{u}_{s^{\prime }}^b(%
\mathbf{r})\widehat{d}_s^c(\mathbf{r})\,M_{s^{\prime }\,s}^\Gamma ,
\end{eqnarray}
where 
\begin{eqnarray}
\,M_{s^{\prime }\,s}^\Gamma &=&\mathit{\upsilon }_{s^{\prime }}^{\top
}C\Gamma \mathit{\upsilon }_s,\, \\
\mathit{\upsilon }_s &=&(\chi _s,0,0), \\
\chi _{\uparrow } &=&(1,0),\,\chi _{\downarrow }=(0,1).
\end{eqnarray}

It is immediate to verify that $M_{s^{\prime }s}^{\gamma _4}=
M_{s^{\prime}s}^1=0$ and therefore the densities $\rho_1({ \bf r})$ and 
$\rho_{\gamma _4}({ \bf r})$ vanish
in the non-relativistic limit. On the other hand, one has 
\begin{eqnarray}
M_{s^{\prime }s}^{\gamma 5} &=&-i\chi _{s^{\prime }}\sigma _2\chi _s \\
M_{s^{\prime }s}^{\gamma 3} &=&\chi _{s^{\prime }}\sigma _1\chi _s \\
M_{s^{\prime }s}^{\gamma _4\gamma 5} &=&i\chi _{s^{\prime }}\sigma _2\chi _s.
\end{eqnarray}
From these relationships it follows that the operators 
$F_{\gamma _5}^a(x)$
and $F_{\gamma _4\gamma _5}^a(x)$ absorb two quarks when they are coupled to
zero total angular momentum, while the operator $F_{\gamma _3}^a(x)$
absorbs them in the $J=1,$ $J_z=0$ channel.

The matrix elements (\ref{rho}) can be computed from the expressions
(\ref{proton})  using
the anti-commutation relations for the quark creation/annihilation
operators.  Alternatively\footnote{We thank D.Diakonov for
making this observation and pointing out an algebraic mistake in the first 
version of this manuscript.}, 
one can obtain them simply by taking the non-relativistic 
limit of the Chiral-Soliton Model results ( i.e. by setting 
$G(\sigma_0 g {\bf r})=0$ 
in Eq.s~(\ref{resSOLpi}-\ref{resSOLpf})~)~\cite{Diakonfock}.

The result is:
\begin{eqnarray}
\rho_{1}(\mathbf{r}) &=& 0,\\
\rho_{\gamma_5}(\mathbf{r}) &=& N_c(N_c-1) 
\int d^3 \mathbf{r'}\psi ^{*}(\mathbf{r',r,r)}\psi (\mathbf{r',r,r)},\\
\rho_{\gamma_4}(\mathbf{r}) &=& 0,\\ 
\rho_{\gamma _4\gamma _5}(\mathbf{r}) &=&  N_c(N_c-1)~
\int d^3 \mathbf{r'}\psi ^{*}(\mathbf{r',r,r)}\psi (\mathbf{r',r,r)},\\
\rho_{\gamma_3}(\mathbf{r}) &=& \frac{N_c (N_c-1)}{3}\int d^3
\mathbf{r'}\psi ^{*}(\mathbf{r',r,r)}\psi (\mathbf{r',r,r)}.
\end{eqnarray}

Notice that, as long as one is interested in ratios of densities, it 
is not necessary to specify the spatial wave-function 
$\psi({\bf r}_1,{\bf r}_2,{\bf r}_3)$.
In the  simple non-relativistic model, 
ratios of diquark densities are completely determined by the color, spin 
and flavor structure of the 
$SU(6)$  nucleon wave-function. This is in general
not the case in  relativistic models, or in $SU(6)$ breaking Non-Relativistic
Quark Models, where spin and spatial degrees of 
freedom do not factorize.

\subsection{Random Instanton Liquid Model}
\label{RILMcalc}

Instantons are topological 
gauge configurations which extremize the Euclidean Yang-Mills action and 
therefore appear as saddle points in the semi-classical approximation to QCD.
They generate the so-called 't~Hooft 
effective quark-quark interaction,
that solves -at least on a qualitative level-
the U(1) problem~\cite{'thooft} and spontaneously breaks 
chiral symmetry \cite{diakonov}. On the other hand, present 
instanton models do not lead to an area-law for the Wilson loop. 

The Instanton Liquid Model  assumes that the QCD vacuum is saturated by 
an ensemble of instantons and anti-instantons.
The only phenomenological parameters in the model are the average instanton
size and density:
\be 
\label{parameters}
\bar{\rho}= \langle \rho\rangle~\simeq~1/3~\textrm{fm},\qquad 
\bar{n}=\langle n\rangle~\simeq~1~\textrm{fm}^{-4}.
\ee
 These values are fixed to reproduce the global 
vacuum properties (quark and gluon condensates)~\cite{shuryak82}.

The Instanton Liquid Model can be used to account numerically 
for the 't~Hooft interaction to all orders.
Such calculations are 
performed by exploiting the analogy between the Euclidean generating 
functional and the partition function of a statistical 
ensemble~\cite{shuryakrev}, in close 
analogy with what is usually done in lattice simulations.
After the integral over the fermionic degrees of freedom is carried out
explicitly, one computes
expectation values of the resulting Wick contractions by
performing a Monte Carlo average over the configurations of an ensemble of
instantons and anti-instantons. 
In each instanton background configuration, the quark propagators are
obtained by inverting numerically the Dirac operator.
In the Random Instanton Liquid, 
the density and the size of the pseudo-particles are kept fixed and coincide
with the
average values (\ref{parameters}). 
On the other hand, the  position and the  
color orientation of each instanton and anti-instanton
are generated according to a random distribution. 

In order to compute the diquark densities in the Random Instanton Liquid Model 
we start by considering 
the following  Euclidean three-point correlation function:
\be
\label{threept}
G^{\Gamma}(x_i,x_f,y)~=~
\langle 0|
~J^\alpha (x_f)\,D_{\Gamma}(y)\, 
\bar{J}^\alpha(x_i)~|0\rangle,
\ee
where $D_\Gamma(y)$ is the diquark density operator defined in (\ref{D}), and
\be
\label{overlappingJ}
J^\alpha (x)= \epsilon^{abc}\,u^{a\,T}(x)\,C\,\gamma_5\,d^b(x)\,
u^{c\,\alpha}(x),\qquad \alpha=1,..,4
\ee 
is an  interpolating operator which 
excites states with the quantum numbers of the proton.
The correlator (\ref{threept}) represents the probability
amplitude to create a state with the 
quantum numbers of a proton at point $x_i$,
to absorb and re-emit two quarks in a given diquark configuration at a the 
point $y$, and to finally  re-absorb the three-quark state at the point~$x_f$.

By inserting two complete sets of eigenstates of the QCD Hamiltonian 
in (\ref{threept}) we obtain:
\be
\label{spectral}
G^{\Gamma}(x_i,x_f,y) =
\sum_{s, s'}
\int \frac{d^3\, {\bf p'}}{2\,\omega_{p'} \, (2\pi)^3}
\int \frac{d^3\, {\bf p}}{2\,\omega_{p} \, (2\pi)^3}
\cdot\,{\textrm Tr}~
[ \la 0| \,J(x_f)\,|\, N({\bf p}',s')\,\ra\nonumber\\
\cdot\,\la N({\bf p}',s')\,|\,D_\Gamma(y)\,|\,N({\bf p},s)\,\ra\cdot
\la\,N({\bf p},s)|\, \bar{J}(x_i)\,|\,0\,\ra] +...,
\ee
where
$|N({\bf p},s)\ra$ denotes a proton state of momentum ${\bf p}$ 
and spin $s$ and 
the ellipses represent all terms depending on  
its excitations (including the continuum contribution).
In the limit of large Euclidean separations (~$|x_i-y|,|x_i-x_f|$ and 
$|x_f-y|~\to~\infty$),
the contribution from the excited states to the correlation function
is exponentially suppressed 
and only the proton state propagates between the operators. 

The overlap of the current operator with the nucleon can be written as:
\be
\label{lambda}
\la 0| \,J(x_f)\,|\, N({\bf p}',s')\,\ra =~
\Lambda 
~\mathit{\upsilon }_s({\bf p})~ e^{i p\cdot x_f},
\ee
where $\mathit{\upsilon }_s({\bf p})$ denotes a Dirac spinor.
Following~\cite{stech2}, the matrix elements of the diquark 
density operator can be parametrized as:
\be
\label{rhotilde}
\la N({\bf p'}, s)| D_\Gamma(y)| N({\bf p}, s)\ra=
h_\Gamma(q^2)~e^{-i q\cdot y}
~\overline{\mathit{\upsilon }}_{s'}({\bf p'})
~\mathit{\upsilon }_s({\bf p}).
\ee

\begin{figure}
\includegraphics[scale=0.5]{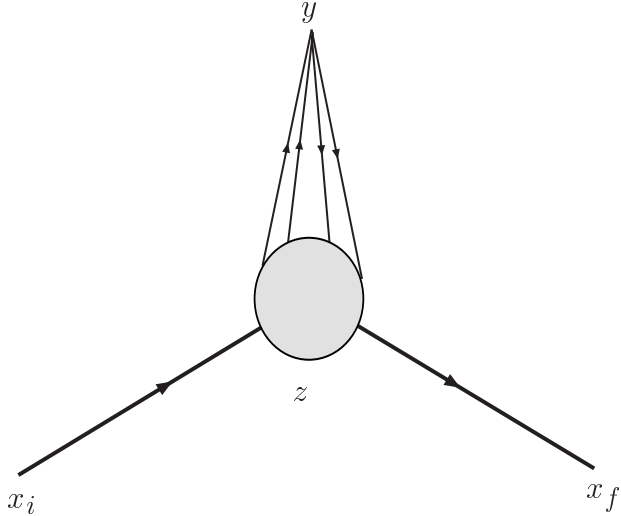}
\caption{Graphical representation of the integrand in 
Eq.~(\ref{ampl}).}
\label{drawfig}
\end{figure}

Substituting (\ref{rhotilde}) and (\ref{lambda}) into (\ref{spectral})
one obtains:
\be
G^{\Gamma}(x_i,x_f,y) &=& \Lambda^2\,
\int \frac{d^3\, {\bf p'}}{2\,\omega_{p} \, (2\pi)^3}
\int \frac{d^3\, {\bf p}}{2\,\omega_{p'} \, (2\pi)^3}
~e^{i ( x_f\cdot p' -x_i \cdot p - y\cdot q) }
\nonumber\\
&\cdot&
~h_\Gamma(q^2)
~\textrm{Tr}~\left[
(\ds{p}'+ M)~(\ds{p} + M)\right]+...
\ee
Next, we use the definition of the free fermion 
propagator in the forward time direction:
\be
\label{Sproton}
S(x';x) =\int\,\frac{d^3 {\bf p}}{(2\pi)^3 \, 2 \, \omega_p}\,
 e^{i\,p \cdot (x'-x)}( \ds{p} + M),\qquad( x'_4 > x_4)\nonumber\\
= \gamma_\mu~(x'-x)_\mu~\Pi_1(|x'-x|) + M~\Pi_2(|x'-x|)
\ee
where
\be
\Pi_1(x)&=&\frac{-i\,M^2}{4\,\pi^2~x^2}~
\left(
K_0(M~|x|)+\frac{2}{M~|x|}~K_1(M~|x|)
\right)\\
\Pi_2(x)&=&\frac{i\,M^2}{4\,\pi^2~|x|}~K_1(M~|x|)
\ee
and introduce the Fourier transform of the function $h_\Gamma(q^2)$,
\be
R_\Gamma(z)=\int \frac{d^4 q}{(2\,\pi)^4} ~e^{i\,z\cdot q} ~h_\Gamma(q^2),
\ee
We obtain:
\be
\label{ampl}
G^{\Gamma}(x_i,x_f,y) = \Lambda^2\,\int d^4 z
~R_\Gamma(y-z)~\textrm{Tr}\,
[~S(x_f;z)~S(z;x_i)~].
\ee
The physical interpretation of this result is the following 
(see also Fig.~\ref{drawfig}).
In the large Euclidean separation limit, the 
correlator $G^\Gamma(x_i,x_f,y)$  is governed by the 
function $R_\Gamma(x)$, which encode information about the probability
amplitude to find the diquark at a given distance 
from the center of the nucleon. Notice that $x_i,x_f,z$ and $y$ 
are four-dimensional
vectors, so Eq.~(\ref{ampl}) accounts for 
relativistic 
retardation effects. On the other hand, the convolution of $R_\Gamma(x)$
with the trace of proton 
propagators takes into account the center of mass motion.

In order to clarify the relationship between 
the correlator (\ref{threept}) and the diquark density (\ref{rho}) 
it is instructive to consider first the static approximation for the nucleon,
in which $M\to~\infty$. In this limit, the proton propagator reads:
\be
\label{staticS}
S(x;y) \simeq \delta({\bf x-y})~e^{-M (x_4-y_4)}~\frac{(1+\gamma_4)}{2},
\ee
and Eq.~(\ref{ampl}) becomes:
\be
\label{easyNR}
G^{\Gamma}(x_i,x_f,y)& =& 2~\Lambda^2\,e^{-(x_f^4-x_i^4)\,M}
~\delta({\bf x_f-x_i})~\int d z_4 ~R_\Gamma({\bf r},z_4),
\ee
where we have used translational invariance to set $y_4=0$ and
we have introduced ${\bf r}:={\bf y}-{\bf x_i}$, the distance 
between the center of the proton and the position 
where the diquark is absorbed.
The last integral term in this expression
represents the time-integrated probability amplitude
to find a diquark at a distance ${\bf r}$ from the center of the nucleon.
We can therefore identify this quantity with the diquark density (\ref{rho})
defined in section~\ref{densities}:
\be
\rho_\Gamma({\bf r})= ~\int d z_4 ~R_\Gamma({\bf r},z_4).
\ee
Hence, for an infinitely heavy nucleon, the expression relating the 
correlation function (\ref{threept}) to the diquark density is simply:
\be
\label{final}
G^{\Gamma}(x_i,x_f,y)& =& 2~\Lambda^2\,e^{-(x_f^4-x_i^4)\,M}
~\delta({\bf x_f-x_i})~\rho_\Gamma({\bf r}).
\ee
If the nucleon mass is kept finite, there are corrections to Eq.~(\ref{final}) 
arising from replacing the static propagator 
(\ref{staticS}) with the exact 
expression (\ref{Sproton})\footnote{We thank J.W.~Negele for his observations
on this point.}.
As a result, the convolution function  $\textrm{Tr}[S(x_f;z)S(z,x_i)]$ in  
(\ref{ampl}) $-$~which determines the position of the
center of mass of the nucleon~$-$  
is delocalized on a volume which depends on the Euclidean time $\tau$.
We have verified that for a typical mass 
$M\simeq~1$~GeV and a typical Euclidean separation  $\tau\simeq$~1~fm  
the position of the center of mass of the nucleon
is smeared on a volume of radius $\simeq$~0.3~fm, centered around the origin. 

A spatial dependence of the point-to-point propagator is a signal that our
nucleon is not at rest. 
Indeed, in order to isolate completely the zero-momentum component of the nucleon wave-function in our calculation, one would need to use very large Euclidean time separations. However, this would be extremely computationally demanding, because the signal-to-noise ratio for the propagators drop exponentially with the Euclidean time separation. 

From these considerations it follows that one should not  {\it quantitatively} compare the detailed shape of the diquark densities  of the Chiral Soliton Model and Constituent Quark Model with that of the Random Instanton Liquid Model. However we shall see below that, as long as one is interested in ratios of diquark densities, it is still possible to draw important qualitative conclusions about the {\it relative} strentgh of the diquark correlations.
 
In this work, we shall estimate ratios of different diquark densities 
$\rho_{\Gamma'}({\bf r})/\rho_{\Gamma}({\bf r})$
by computing ratios of correlation functions 
$G^{\Gamma'}(x_f,x_i,y)/G^{\Gamma}(x_f,x_i,y)$.
Taking ratios allows to maximally reduce the effects of the
corrections due to center of mass motion. In fact, such corrections
affect in the 
same way both the numerator and the denominator and do not change the
normalization of the correlation function.
We have computed (\ref{threept}) 
choosing  ${\bf x}_i={\bf x_f}=0$, $y=(0,0,r,0)$ and $x^4_f=-x^4_i:=\tau$,
with $\tau=$~1~fm.
Previous analysis in the Random Instanton Liquid Model have shown
that the proton pole is isolated from its excited states for 
$\tau\gtrsim~0.9$~fm~\cite{baryonRILM,mymasses,myformfactors}.
The numerical calculation has been done averaging
over 800 configurations of an
ensemble of 492 instantons of $0.33$~fm size, 
in $V~=~4.5^3~\times~5.4~{\textrm fm}^4$ box with the topology of a torus.
We have used rather large 
current quark masses ($m_u=m_d\simeq~100 MeV$) in order 
to reduce finite volume artifacts.

Let us comment on the dependence of the Random Instanton Liquid Model predictions on the value of the bare (current) quark mass chosen in our simulations. It is well known that quarks propagating in the instanton vacuum acquire an effective mass associated to the breaking of chiral symmetry. More specifically, the position of the pole of the quark propagator in the instanton background is shifted by an amount $M_{eff}(p,m_q)$, which depends on both the momentum of the quark $p$ and of its current mass, $m_q$.

The dependence of the effective mass on the current quark mass $m_q$ has been studied in a number of works, using different approximations~(see e.g.~\cite{mussa} and references therein).  In all such studies it was found that the sum of the effective mass at zero momentum and the current quark mass $M_{const}= m_q + M_{eff}(0,m_q)$ (which can be interpreted as the constituent quark mass) remains of the order $\simeq~350$~MeV for all current masses $m_q \lesssim 200$~MeV.
In other words, in this model the constituent quark mass is rather insensitive on the the value of the current quark mass. So, we do not expect that the predictions of Random Instanton Liquid Model presented in this work would change dramatically if a smaller current quark mass was used.

\subsection{Note on Lattice Calculation of Diquark Densities}

We conclude this section by noting that the calculation scheme 
based on point-to-point correlation functions,
used to compute the diquark densities in the Random Instanton Liquid Model, 
can also be applied to compute the same quantities  in Lattice QCD. 
Conceptually, one needs to replace the average over the configurations
of the instanton ensemble with an average over {\it all} 
lattice configurations, performed in the usual way, i.e. 
by sampling the space of lattice links.
Unlike instanton models, lattice QCD calculations are affected by ultraviolet 
divergences, which  are regularized by the lattice spacing.
Hence, in a lattice computation,  the diquark density operators 
$D_\Gamma({\bf r})$  have to be treated as usual Wilson operators 
and need to be renormalized.
\begin{figure}
\includegraphics[scale=0.5]{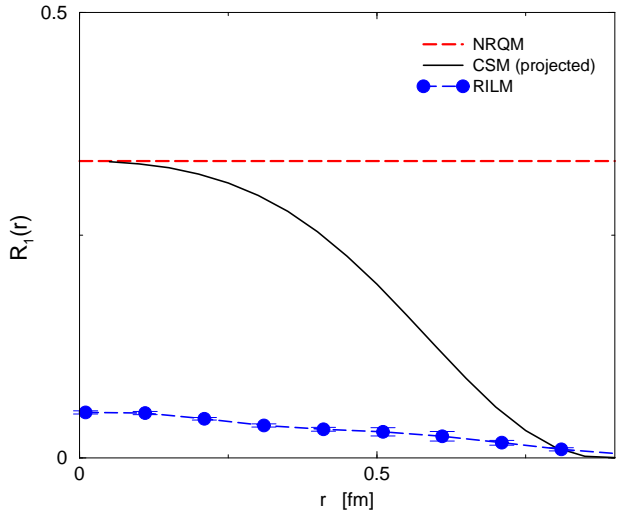}
\caption{Predictions for the ratio 
$R_1(r)=\rho_{\gamma_3}({\bf r})/\rho_{\gamma_5}({\bf r})$ in the 
Non-Relativistic Quark Model (NRQM), in the Random Instanton Liquid Model 
(RILM) and in the Chiral Soliton Model (CSM). The CSM results include the
projection onto the proton state.}
\label{ratios35}
\end{figure}

\section{Comparison and Discussion}
\label{discussion}

In this session, we discuss the results of the phenomenological calculations
presented in the previous session. 
We choose to consider  only the ratios constructed 
by dividing the densities $\rho_{\gamma_3}({\bf r})$. 
$\rho_{\gamma_4\,\gamma_5}({\bf r})$ and
$\rho_{\gamma_4}({\bf r})$ by the scalar density $\rho_{\gamma_5}({\bf r})$. 
There are several reasons for this choice.
In general, ratios are less model dependent than the individual
densities.  For example, in the Non-Relativistic Quark Model, they are 
completely insensitive to
the details of the spatial wave function. Moreover, in lattice and Random 
Instanton Liquid Model
calculations, taking ratios allow to remove
the dependence on the Proton Mass, Euclidean time and the coupling 
$\Lambda$~(see Eq.~(\ref{final})).
We shall not discuss 
the ratio constructed with the $\rho_1({\bf r})$ density, as it is
identically zero in all the models we have considered.

Let us first analyze the ratio constructed with the
density of axial-vector and scalar diquarks: 
\be
\label{R1}
R_1({\bf r})=\frac{\rho_{\gamma_3}({\bf r})}{\rho_{\gamma_5}({\bf r})}.
\ee
The results of our phenomenological calculations 
 are shown in Fig.~\ref{ratios35}.
In the $SU(6)$ Non-Relativistic Quark Model, 
this ratio is completely determined  by the $SU(6)$ spin-flavor
structure of the wave-function, and is identically equal to 1/3.
In the Random Instanton Liquid Model, $R_1({\bf r})$
is sizebly reduced in magnitude (by a factor $\simeq~5$). 
In the Chiral Soliton Model, at ${\bf r}=0$
the $p$-wave contribution from the lower components of the spinors vanishes and one recovers the Non-Relativistic Quark Model Results.
On the other hand, the axial-vector diquark density 
drops down very rapidly at the border of the soliton, where the pion field is most intense.

These results can be interpreted as follows.
In the Random Instanton Liquid Model, the spin- and flavor- dependent 
't~Hooft interaction generates a strong attraction which enhances
the probability amplitude of finding two quarks in the 
same point in the $0^+$ anti-triplet 
configuration, relative to the amplitude of 
finding them in the $1^+$ configuration.  This explains why
the Random Instanton Liquid Model prediction for $R_1({\bf r})$ is 
much smaller than that of the $SU(6)$ Non-Relativistic Quark Model and
the Chiral Soliton Model.
The discrepancy between Random Instanton Liquid Model and mean-field 
Chiral Soliton Model calculations is a signal that in the instanton vacuum at
$N_c=3$ there are strong two-body  correlations, which are not captured by
the large $N_c$ approximation.

From this comparison it follows that a lattice calculation  
of $R_1({\bf r})$ could provide information about the {\it strength} of scalar 
diquark correlations in the nucleon. 
If the non-perturbative QCD interactions generate a 
strong correlation in the $0^+$ anti-triplet channel, as assumed in the
Jaffe-Wilczek model, then we predict that the curve obtained from a 
lattice calculation should lie much below 1/3. 
If lattice simulations found that $R_1({\bf r})\sim 1/3$, 
then this would imply
that  diquarks are not particularly correlated  in the $0^+$
diquark channel and therefore the Jaffe-Wilczek picture is not correct.
$R_1({\bf r})>1/3$ would represent an indication that the quark-quark
interaction is less attractive in the $0^+$ channel, relative to
the $1^+$ channel. This would certainly be a very surprising result, since
the ($\bar{{\bf 3}}_c,~\bar{{\bf 3}}_f$) channel
is favored by both the perturbative and the instanton-mediated interactions.

Unfortunately, the ratio $R_1({\bf r})$ does not encode
information about the microscopic dynamical mechanism underlying such diquark
correlations. In fact, two completely different quark-quark
effective interactions (e.g. one with a chirality-conserving
vertex and one with a chirality-flipping vertex)
may lead to the same predictions, as long as the short-range 
attraction in the scalar channel is sufficiently strong.

In order to gain some insight 
on the microscopic origin of diquarks we need to 
analyze a different ratio:
\be
\label{R2}
R_2({\bf r})= \frac{\rho_{\gamma_5\,\gamma_4}({\bf r})}
{\rho_{\gamma_5}({\bf r})}.
\ee

\begin{figure}
\includegraphics[scale=0.5]{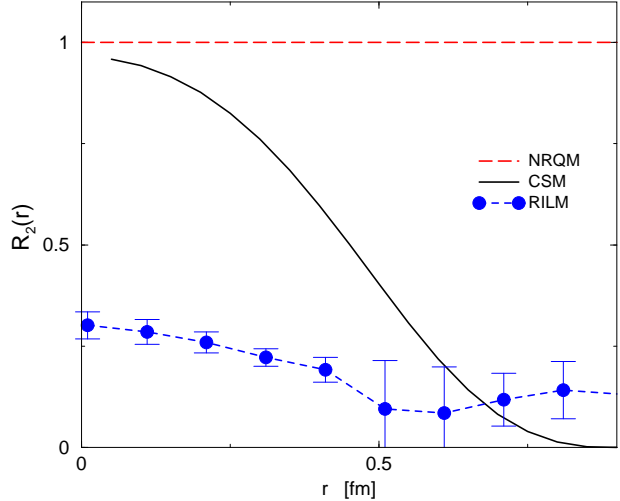}
\caption{Predictions for the ratio 
$R_2({\bf r})=\rho_{\gamma_4\gamma_5}({\bf r})/\rho_{\gamma_5}({\bf r})$,  
in the Non-Relativistic Quark Model (NRQM), in the Random Instanton 
Liquid Model (RILM) and in the Chiral Soliton Model (CSM).}
\label{ratios455}
\end{figure}

The results of our calculations in the three phenomenological
models are reported in  Fig.~\ref{ratios455}.
In the Non-Relativistic Quark Model, both the $\rho_{\gamma_5}({\bf r})$ 
and $\rho_{\gamma_5\,\gamma_4}({\bf r})$
densities probe the $0^+$ scalar diquark content of the proton, so 
 $R_2({\bf r})=1$.
In the Random Instanton Liquid Model the  magnitude of this ratio is smaller than in the Non-Relativistic Quark Model, by a factor 3 or so. 
The fact that, in the Random Instanton Liquid Model, $R_2({\bf r})\ll1$ 
has an important dynamical 
explanation.  It is due to the different sensitivity of the numerator 
and denominator to the so-called {\it direct-instanton} contribution.
The  $\rho_{\gamma_5}({\bf r})$  diquark
density receives maximal contribution from the interaction of quarks
with the field of the closest (direct) instanton, in the vacuum. This
statement can be verified by computing the correlator in the single-instanton 
approximation, discussed in \cite{SIA}. 
On the other hand, the density in the 
numerator, $\rho_{\gamma_5\,\gamma_4}({\bf r})$
 does not receive such 
a direct-instanton contribution and instanton-induced effects come only
from the interactions of quarks with many instantons. 
The magnitude of the 
latter contributions are parametrically suppressed by the 
diluteness of the instanton vacuum, $\kappa=\bar{n}\,\bar{\rho}^4\ll~1$.
Physically, this is the same reason why vector and axial-vector channels have
a rather large ``Zweig rule'', forbidding flavor mixing, while for scalar and
pseudo-scalar channels such a mixing is very strong.

We remark that the very
strong channel dependence of hadronic correlation functions
is a well-known dynamical implication of instanton models. It is
 quite hard to obtain this effect
in alternative dynamical mechanisms~(for a detailed 
discussion see \cite{shuryakcorr}).
This was initially pointed out by
Novikov, Shifman, Vainstein and Zakharov, in the context of Operator
Product Expansion (OPE) and QCD sum-rules for hadronic correlation 
functions~\cite{areallhadronsalike}. Diquark correlations induced by the direct
instantons were discussed also in~\cite{kochelev1}.

In the Chiral Soliton Model, the ratio $R_2({\bf r})$ remains of order 
1 for $|{\bf r}|\lesssim~1$~fm and drops rapidly at the border of the soliton.
The significant
deviation of the Chiral Soliton Model result from the Random Instanton 
Liquid Model prediction
shows that a mean-field approach does not capture correlations
associated to the direct-instanton effects.

From this discussion it follows that, if the scalar diquark
correlations are mainly induced by instantons or in general 
by a Nambu-Jona-Lasinio type of interaction (i.e. chirally symmetric, 
with a chirality flipping vertex), then a lattice measurement should give 
$\rho_{\gamma_4\gamma_5}({\bf r})\ll\rho_{\gamma_5}({\bf r})$, 
so $R_2({\bf r})\ll~1$.

\begin{figure}
\includegraphics[scale=0.5]{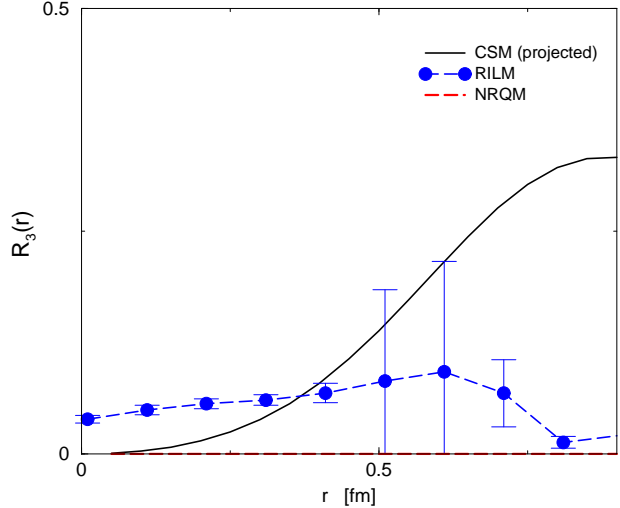}
\caption{Predictions for the ratio 
$R_3({\bf r})=\rho_{\gamma_4}({\bf r})/\rho_{\gamma_5}({\bf r})$ 
in the 
Non-Relativistic Quark Model (NRQM), in the Random Instanton Liquid Model 
(RILM) and in the Chiral Soliton Model (CSM). The CSM results include the
projection onto the proton state.
The Non-Relativistic Quark Model curve is identically zero.}
\label{ratios45}
\end{figure}

Let us now analyze a third ratio: 
\be
\label{R3}
R_3({\bf r})= \frac{\rho_{\gamma_4}({\bf r})}{\rho_{\gamma_5}({\bf r})}.
\ee 
The results of our calculations in the three phenomenological
models are reported in Fig.~\ref{ratios45}.
This quantity is identically zero in the Non-Relativistic Quark Model, 
so it represents a 
probe of relativistic effects in the nucleon.
The ratio is also rather small in the Random Instanton Liquid Model, 
denoting that quarks propagating
in the nucleon receive small relativistic corrections\footnote{We recall that
in both the Random Instanton Liquid Model and in 
the Chiral Soliton Model 
relativistic effects are included.}. 
On the contrary, $R_3({\bf r})$  in the Chiral Soliton Model 
increases rapidly and approaches 1/3, near the border of the soliton.
The interpretation of this result
is the following\footnote{We thank W.~Weise for his comments on this
point.}. 
Near the edge of the soliton,  the quark field
changes very  rapidly. This rapid change gives rise to a large derivative
of the wave-function in coordinate space or, equivalently, to large 
high-{\bf p} components, in momentum space. Such modes 
enhance the contribution from the lower components of the 
spinor wave-function, which give rise to relativistic corrections.

\section{Physical Properties of the Scalar Diquark in the Instanton Vacuum}
\label{sizemass}

In the previous section we have shown that
instantons generate a strong attraction between  a  $u$ and a $d$ 
quark in the color anti-triplet $0^+$ channel.
In this section, we provide unambiguous evidence 
that such forces are strong enough
to form a bound diquark state. We also estimate the diquark size 
by computing its electric charge radius in  
the Random Instanton Liquid Model.

\subsection{Mass of the scalar diquark}

The question if instanton models
lead to a bound $0^+$ color anti-triplet 
diquark has been first posed by Diakonov and Petrov~\cite{petrov} and
investigated in a number of works. 
In an exploratory study~\cite{baryonRILM} Shuryak, Sch\"afer and Verbaarshot
computed some diquark two-point functions 
in coordinate space, using the Random Instanton Liquid Model.
They found some evidence for a light diquark bound-state 
(of rougly 450~MeV mass), by performing an analysis of the 
correlation function, based on a 
pole-plus-continuum parametrization of the spectral density.
On the other hand, Diakonov, Petrov and collaborators
have analyzed diquarks by
solving Schwinger-Dyson equations at the leading order in
$1/N_c$~\cite{diak}. 
They found evidence for correlations, but no 
binding\footnote{Except for $N_c=2$, in which case
the diquark is a baryon and its 
mass is protected by Pauli-Gursey symmetry.}.

In order to clarify this issue, we have followed an 
approach which is usually applied to extract hadron masses in
lattice simulations. As usual, we have replaced the average over all gauge 
configurations
with an average over the configurations of the instanton ensemble.
This method presents some 
advantages with respect to the coordinate-representation calculation
of Shuryak {\it et al}. 
On the one hand, it avoids 
the undesired additional model dependence associated with the
parametrization of the spectral function.
On the other hand, it makes unambiguously evident the existence of the bound 
state and allows to determine more precisely its mass.

The starting point consists of computing the Euclidean two-point function:
\be
\label{G2}
G2(\tau)=\int d^3 {\bf r} 
~\langle 0|~T[~J_D({\bf 0},\tau)J^\dagger_D({\bf r},0)\,
Pe^{\int dy_\mu A_\mu(y)}~]|0\rangle,
\ee
where $J_D(x)$ is the usual diquark current:
\be 
\label{J_D}
J^a_D(x)= \epsilon^{abc}\,u^{b\,T}(x)\,C\,\gamma_5\,d^c (x).
\ee
The path-ordered exponent in (\ref{G2})
represents a Wilson line
connecting the two extremes of the two-point function and 
is needed to assure gauge invariance of the correlator. 
In the instanton vacuum,
the contribution from the Wilson line is very small, as 
heavy quarks couple very weakly with instantons.

In the large Euclidean time limit, 
only the lightest state with the quantum numbers of 
the current $J_D(x)$ propagates 
in the two-point function. If there is a diquark bound-state, 
then, in the large Euclidean time limit,
 the logarithm of the two-point function must scale linearly with $\tau$:
\be
\ln G2(\tau)\stackrel{\tau\to\infty}{\rightarrow} \ln \Lambda_D^2 -\tau~M_D,
\ee
where $M_D$ is the mass of the diquark and the constant 
$\Lambda_D$ is its coupling to the current, 
defined by  $\langle 0|J_D(0)|D\rangle=~\Lambda_D$.

\begin{figure}
\includegraphics[scale=.4, angle=-90]{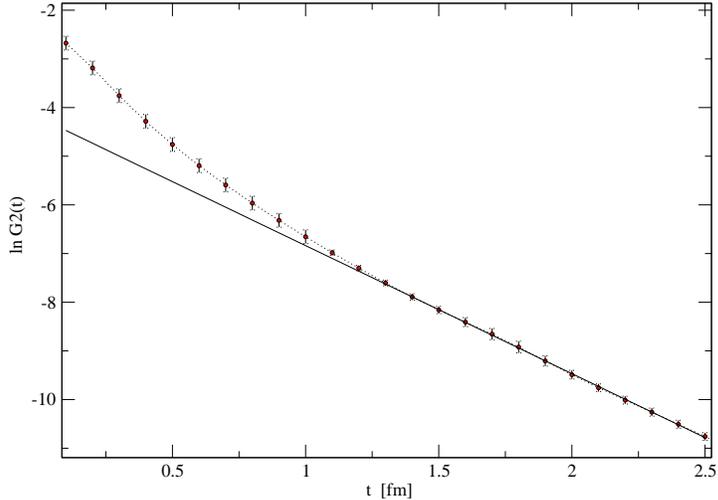}
\caption{Logarithm of the diquark two-point function, $\ln G2(\tau)$ computed
in the Random Instanton Liquid Model. The linear slope is a clean 
signature of the existence of a bound state.}
\label{lnplot}
\end{figure}

We have computed such a correlation function in the Random Instanton 
Liquid Model, averaging over
100 configurations in a $4^3~\times~8~\textrm{fm}^4$ box.
In analogy with lattice simulations, a rather large quark mass 
was needed to reduce the finite-volume artifacts\footnote{We 
choose quark masses of the same order of magnitude of 
the smallest bare masses used in present lattice simulations.}.  
The result for our calculation of the correlation function (\ref{G2}) 
with $m_q\simeq~100$~MeV is shown in Fig.~\ref{lnplot}.
We clearly see that, in the  large $\tau$ limit,
the logarithm of the two-point function
becomes a linear function of $\tau$.

\begin{figure}
\includegraphics[scale=0.5, angle=-90]{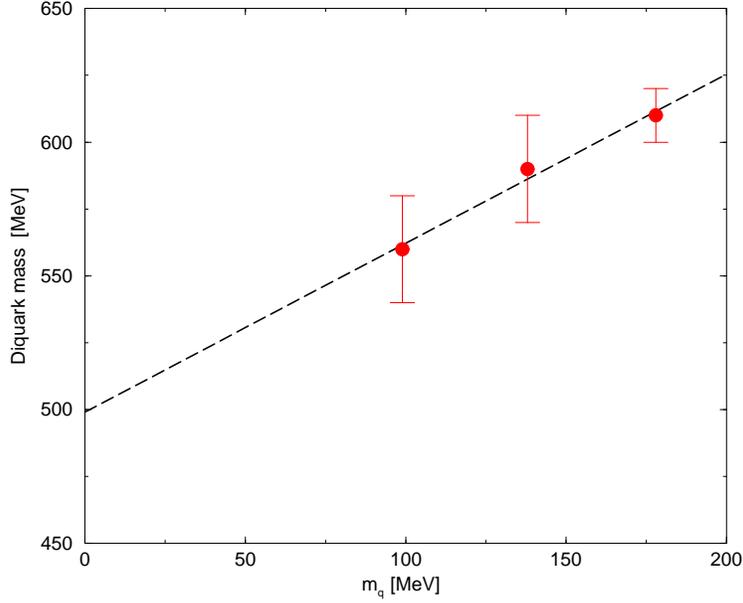}
\caption{The mass of the u-d diquark, computed in the 
Random Instanton Liquid Model,  as a function of 
the $u$ and $d$ bare quark mass.}
\label{figmasses}
\end{figure}

In order to study the sensitivity of the diquark
mass on the current quark mass, we have computed the correlation function
for different values of the $u$ and $d$ quark mass, 
$m_u=m_d=m_q=\{180,~140,~100\}~$MeV.
The dependence of diquarks mass on the current quark masses is plotted in 
Fig.~\ref{figmasses}. 
A rough estimate of the diquark mass can
be obtained by linearly 
extrapolating  to the physical value of the current quark 
mass\footnote{A precise extrapolation fit should include logarithmic
corrections associated with chiral physics. We do not need to discuss these 
effects here, as we are interested in the order of magnitude of the diquark
mass.}. 
We find a diquark mass of $\sim~500$~MeV, in good agreement with 
the previous estimate. 

\subsection{Diquark size}

An important question to address is the size of the diquark.
From general arguments, we expect that it should be 
comparable with the size of the proton. In fact 
even the most tightly 
bound QCD excitation, the pion, is known to have
a rather large electric charge square radius, $r_E^\pi\simeq~0.6$~fm.
On the other hand, an early naive analysis in the Instanton Liquid 
Model~\cite{3ptILM} gave a surprisingly small estimate for the diquark
charge radius,
$r_E\sim~0.3$~fm. It is hard to estimate the accuracy of such an estimate 
which was based on a number of assumptions\footnote{For example,
it was assumed that the diquark form factor follows a monopole fit}.
Hence, it is worth performing
a direct calculation of the diquark electro-magnetic form factor.

To this end, we have
computed the diquark electro-magnetic three-point function,
\be
\label{G3D}
G3_D(\tau,{\bf q})=\int d^3 {\bf y} \,d^3 {\bf x}
~e^{i~{\bf q} \cdot {\bf y}}~\langle 0|~T[~J_D({\bf 0},\tau)~
J^{e/m}_4({\bf y},0)~
J^\dagger_D({\bf x},-\tau)~
Pe^{\int dy_\mu A_\mu(y)}~]|0\rangle
\ee
where $J_D(x)$ is the usual diquark interpolating operator and
$J^{e/m}_\mu(x)=2/3~\overline{u}(x)~\gamma_\mu~u(x)
-1/3~\overline{d}(x)~\gamma_\mu~d(x)$ is the electro-magnetic current operator.
In the large $\tau$ limit, the correlation functions (\ref{G3D})
relates directly to the diquark form factor~\footnote{For a 
discussion of these relationships and of the details of the numerical
 integration method used for performing the Fourier Transform in (\ref{G3D})
 see \cite{pionFF,myformfactors}.}, $F_D(Q^2)$:
\be
G3_D(\tau,{\bf q})\stackrel{\tau\to\infty}{\rightarrow}
\Lambda_D^2 ~e^{-(M+\sqrt{M_D^2+{\bf q}^2})\tau}~F_D(Q^2),
\ee
where the values of the 
diquark mass and the constant $\Lambda_D$ can be extracted from the 
two-point correlation function.

\begin{figure}
\includegraphics[scale=0.4,angle=-90]{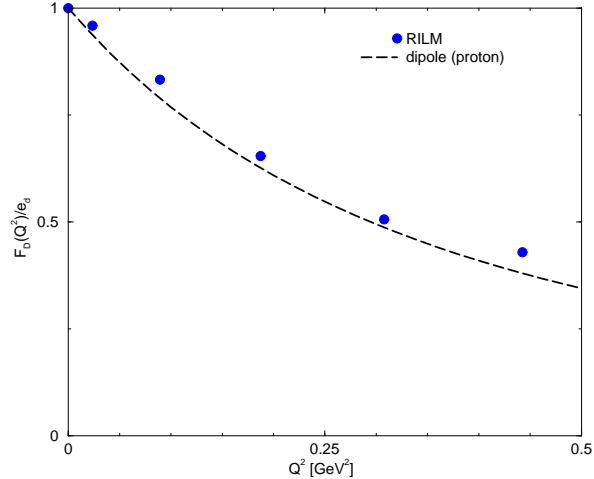}
\caption{The diquark form factor (normalized to the total charge) evaluated
in the Random Instanton Liquid Model (points) 
and compared with the dipole parametrization of the 
low-energy data on proton electric form factor: $F_{dip}=1/(1+Q^2/0.71)^2$ 
(dashed line). 
The numerical simulation was performed in a $(4^3\times 6)~\textrm{fm}^4$ 
box and the exchanged momentum is quantized accordingly.}
\label{G3fig}
\end{figure}

The result of the Random Instanton Liquid Model calculation of the form factor
(normalized to the total $u-d$ diquark charge,~$e_D=1/3$~) is presented 
in Fig.~\ref{G3fig}, where it is compared to the phenomenological 
dipole fit  which reproduces the 
low-energy data on the proton electric form factor.
These results imply that the diquark is an extended object, whose size 
is comparable with that of the nucleon. 
Its electric charge radius is found to be
$r_E^2\textrm{(diquark)}\simeq (0.70~\textrm{fm})^2$, which should be compared to the value
of the electric charge radius of the proton, $r_E^2\textrm{(proton)}\simeq
(0.76~\textrm{fm})^2$ computed in the Random Instanton Liquid Model in 
\cite{myformfactors}.
As expected, the electric charge radius of the diquark is larger than that of the pion  and smaller than that of the proton. 




\section{Conclusions}
\label{conclusions}
In this work we have addressed the question of how is it possible to study
two-body diquark correlations in hadrons, using lattice QCD.
We have identified some suitable  lattice-calculable 
correlation functions, which allow to probe directly the diquark content of 
the nucleon.
We have argued that these Green's functions
provide important information about non-perturbative correlations, 
in different color anti-triplet diquark channels.

In particular, the ratio $R_1({\bf r})$, defined in Eq. (\ref{R1}), 
measures the strength 
of the correlations  in the {\it scalar} diquark channel, relative
to that in the axial-vector channel.
It can be used to check the main dynamical assumption of the Jaffe-Wilczek
 model. 
In fact, if the quark-diquark picture is correct,
then we predict that a lattice measurement must lead to $R_1({\bf r})\ll1/3$.

The ratio $R_2({\bf r})$, defined in Eq. (\ref{R2}),  
can be used to gain some insight on
 the dynamical origin of the non-perturbative interaction
in the scalar diquark channel.  
In particular, it can be used to check the hypothesis according to which 
these forces  are mediated
by instantons (or more generally
by a NJL-like, chirality-flipping interaction).
We have argued that, in this case, we  expect that lattice measurements
should give $R_2({\bf r})\ll1$.

We have computed these ratios using three phenomenological models. 
We have found that they lead to radically different 
predictions for the matrix  elements we have selected. 
Hence, a lattice measurement 
could point out which picture is most realistic.
It is worth stressing 
that from the fact that the Chiral Soliton Model predictions differ 
substantially from the Random Instanton Liquid Model results one 
should not conclude that the Chiral Soliton Model
is not correctly reproducing the physics of instanton-induced interactions.
In fact, such mean-field model can account for one-body local 
operators. On the other hand, our results show that it is much less 
reliable for computing matrix elements of two-body operators.
The discrepancy between Random Instanton Liquid Model and 
Chiral Soliton Model should be taken as 
an indication that rather strong direct two-body correlations are generated in the instanton vacuum.

We would like to stress that the qualitative differences between the predictions of the three models considered in the present analysis are dramatic and therefore quite robust.  
However, one should be careful in comparing the {\it details} of the curves associated to the diquark density ratios, at a quantitative level. 
In fact, on the one hand, the Random Instanton Liquid Model predictions are affected by the smearing effects associated to the center of mass motion (see the discussion in section \ref{RILMcalc}). 
On the other hand, the Chiral Soliton Model  and the Random Instanton Liquid Model calculations are performed using different values~\footnote{In the Chiral Soliton Model calculation we have assumed massless current quarks, while in the Random Instanton Liquid Model calculation we have used current quarks of $m_q\simeq 100$~MeV.} of the current quark mass $m_q$. This fact may reflect itself in a slightly different value of the effective mass of the quarks propagating in the chirally broken vacuum (see the discussion at the end of section \ref{RILMcalc}).

In the second part of the work, we 
have studied in detail the physical properties of instanton-induced 
diquarks. We have provided unambiguous evidence that
instantons  generate a scalar 
diquark bound-state of mass $\sim~500$~MeV, 
in good agreement with earlier estimates based on point-to-point correlation 
functions in coordinate space. 

We have also studied its electric charge distribution.
Our results show that the scalar diquark is an extended object,
whose size is of the order of the fm, hence
 comparable with that of the proton. Thus, phenomenological
quark-diquark models cannot treat the diquark as a point-like object.

As a concluding remark, we stress that the analysis performed in this work 
focused on the diquark content of the proton.
However, the same study could be repeated, in order to investigate the 
diquark content of other lowest-lying baryons. For example, it
 would be particularly interesting
to compare the diquark densities in the nucleon and in the delta.
In fact, in the delta, the leading direct-instanton effects are suppressed,
due to the Dirac- and flavor- structure of the 't~Hooft interaction.
Hence, if diquark correlations are 
instanton-mediated, then we expect that the strong
enhancement of the scalar density observed in the proton
should be much less pronounced in the delta.

\acknowledgments
We would like to thank B.~Golli 
for providing us with the quark wave-functions in the Chiral Soliton Model
and D.~Diakonov for important comments.
The code for computing correlation functions in the Random 
Instanton Liquid Model was 
developed and kindly made available by T.~Schaefer and E.V.~Shuryak.
P.F. acknowledges important discussions with L.~Glozman, J.W.~Negele,
K.~Goeke, R.L.~Jaffe, S.~Simula and W.~Weise.

\end{document}